\documentclass[twocolumn,showpacs,preprintnumbers,amsmath,amssymb,floatfix]{revtex4-1}
\usepackage{graphicx}
\usepackage{dcolumn}
\usepackage{bm}
\usepackage{color}

\begin{document}

\title{High magnetic field evolution of ferroelectricity in CuCrO$_{2}$}

\author{Eundeok Mun$^{1}$, M. Frontzek$^{2}$, A. Podlesnyak$^{3}$, G. Ehlers$^{3}$, S. Barilo$^{4}$, S. V. Shiryaev$^{4}$, Vivien S. Zapf$^{1}$}

\affiliation{$^{1}$National High Magnetic Field Laboratory (NHMFL), MPA-CMMS group, Los Alamos National
Laboratory (LANL), Los Alamos, NM 87545, USA}%
\affiliation{$^{2}$Laboratory for Neutron Scattering, Paul Scherrer Institute, 5232 Villigen-PSI, Switzerland}%
\affiliation{$^{3}$Quantum Condensed Matter Division, Oak Ridge National Laboratory, Oak Ridge, TN 37831-6475, USA}%
\affiliation{$^{4}$Institute of Solid State and Semiconductor Physics, Minsk 220 072, Belarus}%

\begin{abstract}
CuCrO$_{2}$ offers insights into the different types of spiral magnetic orderings that can form spontaneously due to frustration in triangular-lattice antiferromagnets. We explore the
magnetic phase diagram up to 65 T along all the principle axes, and also use electric polarization to probe changes in the spiral order at high magnetic fields. It is known that at
zero magnetic field a proper-screw spiral of the Cr \textbf{S} = 3/2 spins forms that in turn induces electric polarization with six possible orientations \textbf{ab}-plane. Applied
magnetic fields in the (hard) \textbf{ab}-plane have been shown to induce a transition to cycloidal spiral magnetic order above 5.3 T in those domains that have spins perpendicular to
the applied magnetic field. We show that the cycloidal order remains unchanged all the way up to 65 T, which is one quarter of the extrapolated saturation magnetization. On the other
hand for magnetic fields along the (easy) \textbf{c}-axis, we observe a transition in the electric polarization near 45 T, and it is followed by a series of steps and/or oscillations
in the electric polarization. The data is consistent with the a proper-screw-to-cycloidal transition that is pushed from 5.3 to 45 T by easy-axis anisotropy, and is in turn followed
by stretching of the magnetic spiral through commensurate and incommensurate wave vectors. This work also highlights the ability of the magnetically-induced electric polarization to
probe complex magnetic orders in regimes of phase space that are difficult to reach with neutron diffraction.
\end{abstract}


\maketitle

\section{Introduction}

Ferroelectricity can be induced by magnetic ordering that breaks spatial-inversion symmetry (SIS). These magnetic structures are often complex and result from geometrical magnetic
frustration \cite{Kimura2007, Cheong2007}. A well-studied SIS-breaking magnetic structure is the cycloidal magnetic spiral state that has been studied in TbMnO$_{3}$
\cite{Kenzelmann2005} among others  \cite{Takahashi2008, Shuvaev2010, Lautenschlger1993, Taniguchi2008, Shanavas2010, Klimin2003, Kenzelmann2007, Masuda2004, Huvonen2009, Lawes2005,
Harris2006}. The microscopic mechanism in these compounds is thought to be an inverse Dzyaloshinskii-Moriya (DM) \cite{Sergienko2006}, also described as a spin-current model
\cite{Katsura2005}. In these cycloidal spirals, the electric polarization $\textbf{P}$ is perpendicular to the spiral propagation vector $\textbf{q}$ and parallel to the spiral plane:
$\textbf{P} \perp \textbf{q}$. However, recently it has been shown that besides cycloidal spirals, proper screw-type spirals (PSS) can also create magnetic field-induced
ferroelectricity in crystals with certain symmetry \cite{Seki2009, Nakajima2010, Kimura2008}, including the triangular-lattice antiferromagnets (TLAs) CuFeO$_{2}$ with Al- and
Ga-doping and CuCrO$_{2}$. A microscopic theory has been proposed for ferroelectricity induced by a magnetic PSS \cite{Arima2007} where the metal-ligand $d-p$ hybridization is
influenced by spin-orbit coupling. In this mechanism, $\textbf{P}$ is parallel to the screw axis $\textbf{q}$ and perpendicular to the spiral plane: $\textbf{P} \parallel \textbf{q}$.

\begin{figure}
\centering
\includegraphics[width=1\linewidth]{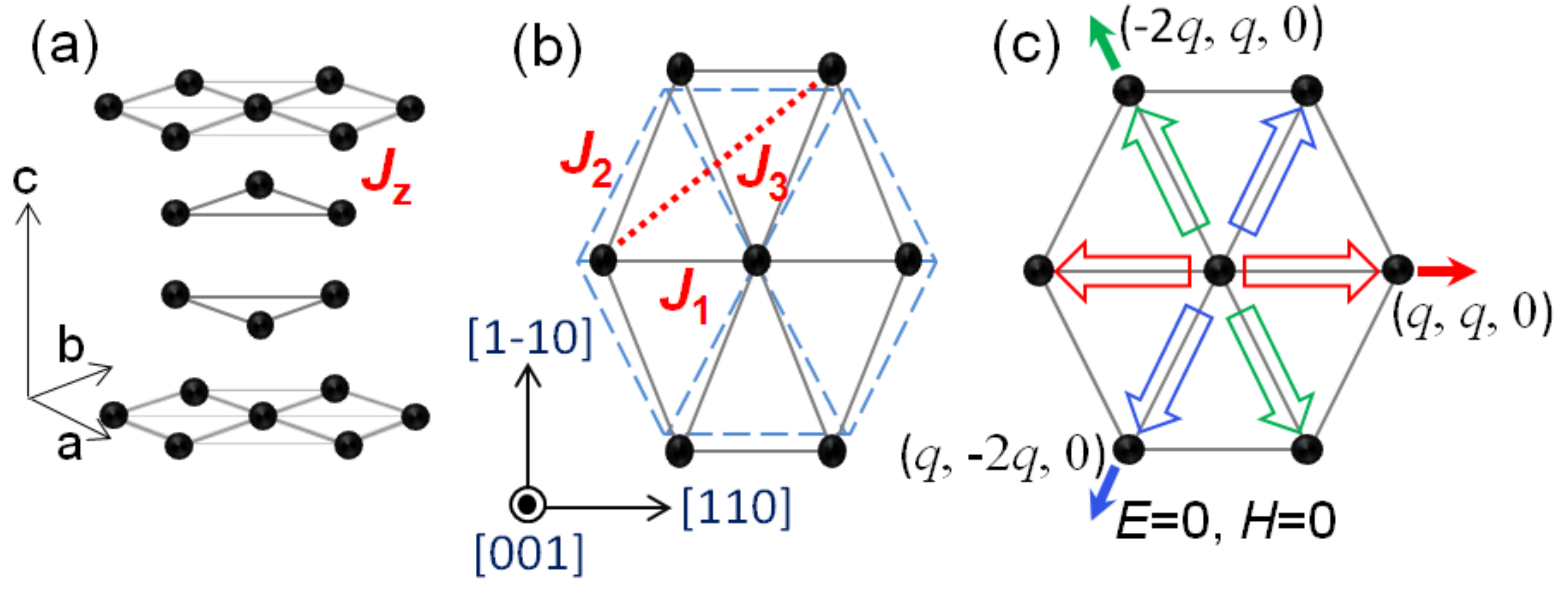}
\caption{(a) Simplified crystal structure of CuCrO$_{2}$ showing the \textbf{S} = 3/2 Cr spins. The inter-plane magnetic exchange interaction $J_z$ is indicated. (b) The
(001)-projection of the Cr triangular lattice plane, where dashed and solid lines represent the hexagonal lattice above and below the magnetic ordering temperature \cite{Kimura2009a}.
The parameters $J_{1}$, $J_{2}$, and $J_{3}$ indicate nearest, next-nearest, and next-next-nearest magnetic exchange interactions within \textbf{ab}-plane. (c) Schematic
representation of ferroelectric domain configuration in zero electric \textbf{E} and magnetic $\textbf{H}$ field. The possible electric polarization directions, which lie along the magnetic spiral propagation vectors of ($\textbf{q}$, $\textbf{q}$, 0),
($\textbf{q}$, -$2q$, 0), and ($2q$, -$\textbf{q}$, 0) are indicated by the different colored arrows. These six directions are all crystallographically equivalent to each other.}
\label{Fig1}%
\end{figure}

The multiferroic CuCrO$_{2}$ compound exhibits both PSS order at zero field, and cycloidal spiral order in applied magnetic fields. It is also an example of a system whose
magneto-electric properties are tunable by both an electric and magnetic field \cite{Kimura2009b, Soda2009a}. This material crystallizes in the rhombohedral $R$\={3}m structure,
forming a triangular lattice of Heisenberg Cr$^{3+}$ (3$d{^3}$ \textbf{S} = 3/2) spins \cite{Kadowaki1990, Crottaza1996, Seki2008}. Along the \textbf{c}-axis, each triangular plane is
rotated from its neighbors by 30 degrees and displaced such that the atoms of one layer reside in the center of triangle of the neighboring layers. The structure of the Cr atoms is
shown in Fig. \ref{Fig1} (a).

Thermodynamic investigations for single crystal samples of CuCrO$_{2}$ reveal that two successive magnetic transitions occur at $T_{N}$ = 24.2 K and $T_{MF}$ = 23.6 K, and nonzero
\textbf{P} appears below $T_{MF}$ \cite{Kimura2008}. The magnetic structure of CuCrO$_{2}$ at zero magnetic field has been identified by neutron diffraction studies as an
incommensurate PSS spin order of the Cr \textbf{S} = 3/2 moments with a propagation vector $q$ = (0.329,0.329, 0) \cite{Poienar2009, Soda2009a, Poienar2010, Soda2010, Kajimoto2010}.
Within the spiral, the ordered moment reaches maximum values of 2.2 $\mu_{B}$ along [1-10] and 2.8 $\mu_{B}$ along [001] \cite{Frontzek2011, Frontzek2012}. Single-crystal neutron
diffraction measurements reveal that the spin structure between $T_{N}$ and $T_{MF}$ (where there is no ferroelectric order) is two-dimensional (2D) with short-range magnetic
correlations along the \textbf{c}-axis \cite{Frontzek2012}. However, in the multiferroic state below $T_{MF}$, the magnetic structure becomes 3D with long-range correlations along
\textbf{c}-axis \cite{Frontzek2012}. Analysis of the magnetic excitation spectrum revealed antiferromagnetic (AFM) nearest- and next-nearest-neighbor exchange coupling within the
plane and further showed that AFM next-next-nearest neighbor exchange couplings $J_{3}$ in the plane also have to be considered \cite{Frontzek2011}. Additionally, it was shown that a
small ferromagnetic (FM) inter-plane coupling ($J_{z}$ $\ll$ $J_{1}$) exists. The inter-plane interaction mediates the 3D-order in the multiferroic state and is responsible for the
incommensurate $\textbf{q}$, deviating from the out-of-plane 120$^{o}$ spin structure ($\textbf{q}$ = 4/3$\pi$\textbf{x}) following the analysis of Ref. \cite{Rastelli1986}. The
relevant exchange interaction paths are shown as a schematic cartoon in Fig. \ref{Fig1} (b). Finally, there is an easy-axis anisotropy $D_z$ along the \textbf{c}-axis.

There is a three-fold rotation symmetry with respect to the hexagonal \textbf{c}-axis, and thus there are six equivalent magnetic spiral domains, since the spiral propagation vector
can point forward or backward along any of the three principle axes \cite{Frontzek2012}. Since the electric polarization points along the spiral propagation vector, there are also six
electric domains as shown in Fig. \ref{Fig1} (c). These domains can be selected by applied electric and/or magnetic fields, and the application of both an electric and a magnetic
field parallel to each other in the \textbf{ab}-plane are proposed to select a single domain \cite{Kimura2009b}. A small lattice distortion was observed in the magnetically ordered
phase by X-ray diffraction and magnetostriction measurements \cite{Kimura2009a}. This distortion changes the equilateral triangles in the \textbf{ab}-plane into isosceles triangles
(Fig. \ref{Fig1} (b)), thus impressing the lowered symmetry of the magnetic order of a given domain onto the lattice \cite{Kimura2009a}. This distortion also induces an anisotropy
$D_x$ in the \textbf{ab}-plane that is selected by the spiral propagation vector. \cite{Frontzek2011}

By applying magnetic fields $\textbf{H} \parallel$ [1-10], the direction of \textbf{P} is changed from [110] to [1-10] at $H_f$ $\sim$ 5.3 T \cite{Kimura2008, Soda2009a, Kimura2009b,
Yamaguchi2010}. Neutron diffraction measurements find that this change in the direction of $\textbf{P}$ is accompanied by a spin flop in the magnetic domain whose spins originally lay
perpendicular to the applied magnetic field. This spin flop corresponds to a transition from a proper-screw-spiral with [110] propagation vector to a cycloidal-spiral spin structure
with a [1-10] propagation vector \cite{Soda2010}. According to the inverse-DM model, the cycloidal spiral could produce a $\textbf{P}$ along [001], however in fact $\textbf{P}$ in the
cycloidal spiral state points along the propagation vector [1-10]. Furthermore, no feature in the magnetostriction is observed at this spin flop transition. Thus no inverse-DM
interaction can be observed in this compound \cite{Kimura2009a} and the multiferroic behavior appears to be dominated by the Arima mechanism \cite{Arima2007}.

In the present study, we extend the phase diagram of CuCrO$_2$ to 65 T for all principle directions of $\textbf{H}$ and $\textbf{P}$ via measurements of magnetization and electric
polarization in pulsed magnetic fields. We uncover additional magnetic and ferroelectric phase transitions.

\section{Experimental}

Single crystals of CuCrO$_{2}$ were grown by a flux method as described previously \cite{Frontzek2012}. The temperature and magnetic field dependence of the magnetization,
$M(T,H)$, were measured in a commercial SQUID (superconducting quantum interference device) magnetometer in a Quantum Design (QD) Magnetic Property Measurement System (MPMS) up to
$\mu_{0}H$ = 7 T and down to $T$ = 2 K. The magnetization data taken in the superconducting magnet are consistent with earlier reports \cite{Kimura2008, Kimura2009b}, where two magnetic
transitions are detected in d$M$/d$T$ at $T_{N}$ = 24.2 K and $T_{MF}$ = 23.6 K and an anomaly is revealed in d$M$/d$H$ curves around 5.3 T along $\textbf{H}$ $\parallel$
[1-10]. In addition, $M(H)$ measurements were extended up to $\mu_{0}H$ = 60 T and down to $T$ = 0.6 K in short-pulsed magnets with an extraction magnetometer
\cite{Detwiler2000} at the National High Magnetic Field Laboratory (NHMFL) at Los Alamos National Laboratory. The magnetization curves taken in pulsed magnetic field are calibrated by the SQUID data.

\begin{figure}
\centering
\includegraphics[width=1\linewidth]{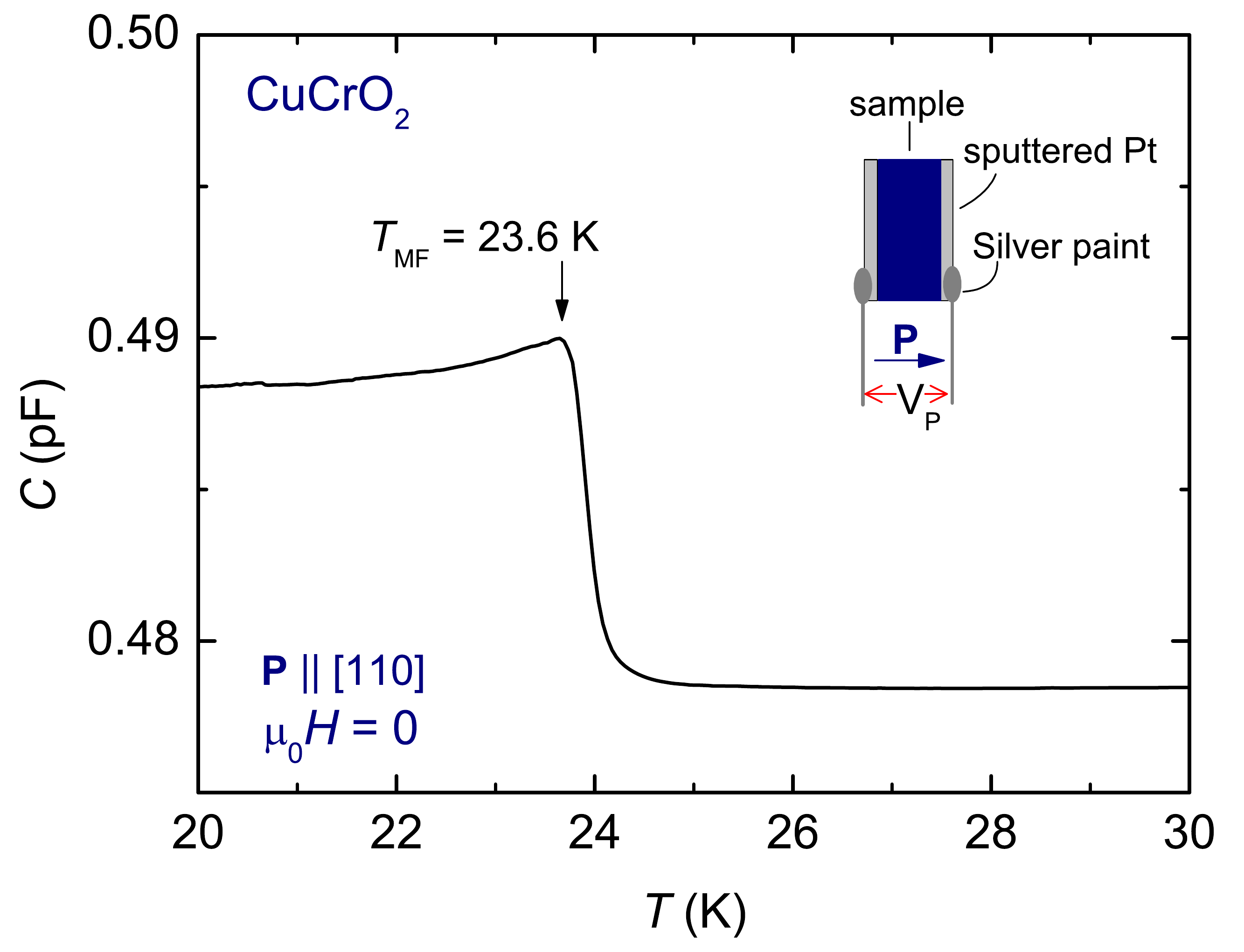}
\caption{Capacitance measurements in zero magnetic field for single crystals of CuCrO$_{2}$. Platinum (Pt) electrodes were sputtered onto the opposite sides of the samples (see
inset). Silver paint was used to make the electrical contact between sputtered Pt electrode and Pt wire. The sample was poled by applying external voltage $V_{p}$ across the contacts.}
\label{Fig2}%
\end{figure}

For the measurements of electric polarization and capacitance, samples were cut into thin plates with dimensions of 0.55$\times$1.17$\times$0.15 mm$^{3}$ for \textbf{P} $\parallel$
[110], 0.58$\times$0.97$\times$0.31 mm$^{3}$ for \textbf{P} $\parallel$ [1-10], and 0.89$\times$1.12$\times$0.38 mm$^{3}$ for \textbf{P} $\parallel$ [001]. The measurements for
\textbf{P} in the \textbf{ab}-plane were consistent across three different  samples.  Platinum electrodes were sputtered onto the samples and then the platinum wires were attached
onto the opposite faces of the thin plate (see the inset of Fig. \ref{Fig2}). The samples were characterized by temperature-dependent capacitance measurements with an AH 2500
Capacitance Bridge at 1 kHz. In zero magnetic field, the phase transitions are seen in capacitance measurement as shown in Fig. \ref{Fig2}, where a distinct peak structure is observed
at $T_{MF}$ = 23.6 K. The magnetic susceptibility measurements show two transitions, $T_{N}$ and $T_{MF}$, whereas capacitance measurements only show the $T_{MF}$ phase transition.

The electric polarizaton in pulsed magnetic fields up to 65 T was measured by integrating the magnetoelectric current induced by changing magnetic fields, recorded using a Stanford
Research 570 current to voltage converter \cite{Zapf2010, Zapf2011}.  The magnetoelectric current is the magnetic field-induced analog of the pyroelectric current, and it increases
linearly with magnetic field sweep rate such that the signal-to-noise of the electric polarization measurements is improved in our pulsed magnets compared to our DC magnets. Electric
polarization was measured after first poling the sample by applying a constant external 200 V across the capacitor plates as the sample is cooled from 30 K to the measurement
temperature. This corresponds to a poling electric field of $E_{P}$ $\sim$ 1.3 MV/m for \textbf{P} $\parallel$ [110] and $\sim$ 0.6 MV/m for \textbf{P} $\parallel$ [1-10] and [001].

\section{Results}

\subsection{Magnetization}

\begin{figure*}
\centering
\includegraphics[width=0.5\linewidth]{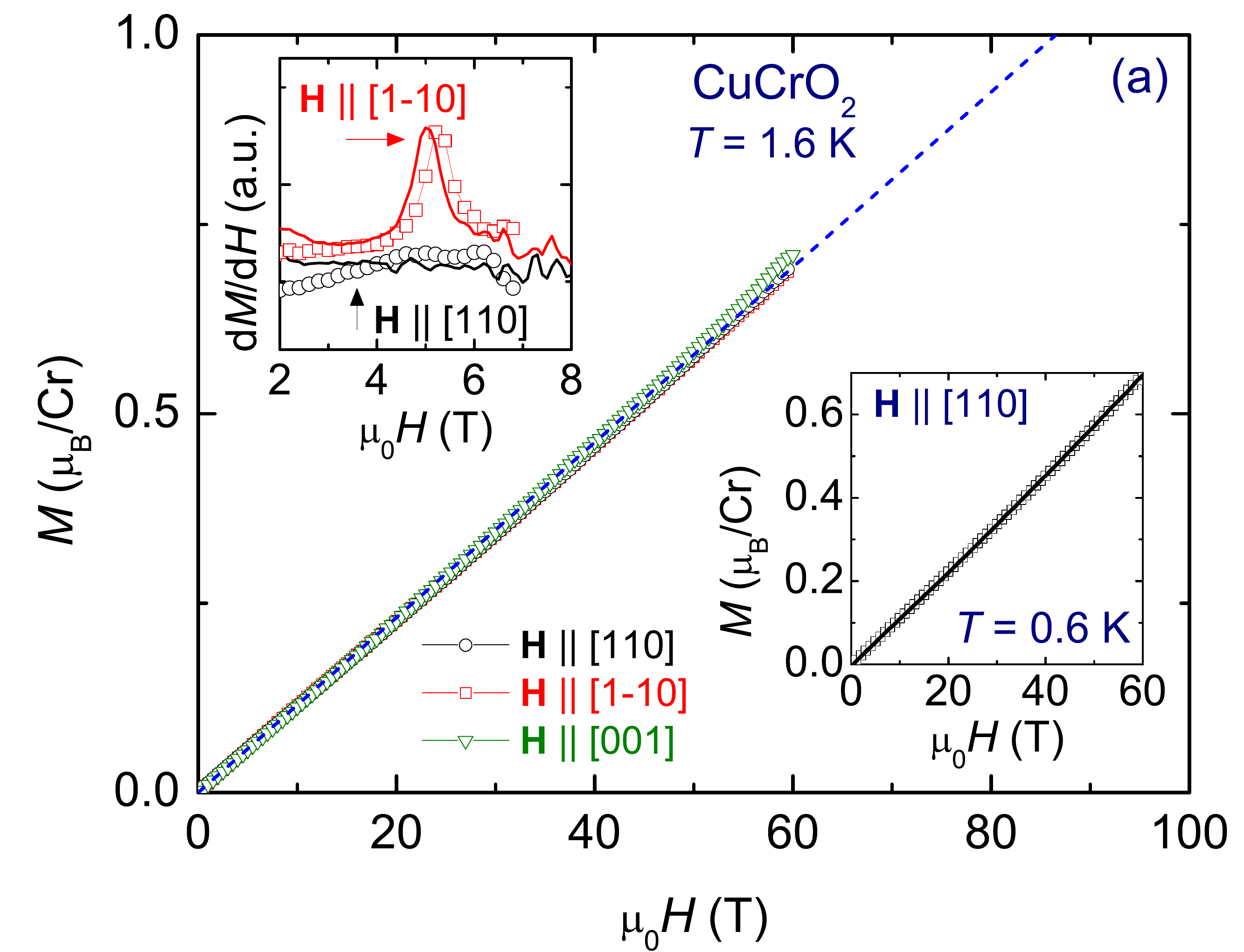}\includegraphics[width=0.5\linewidth]{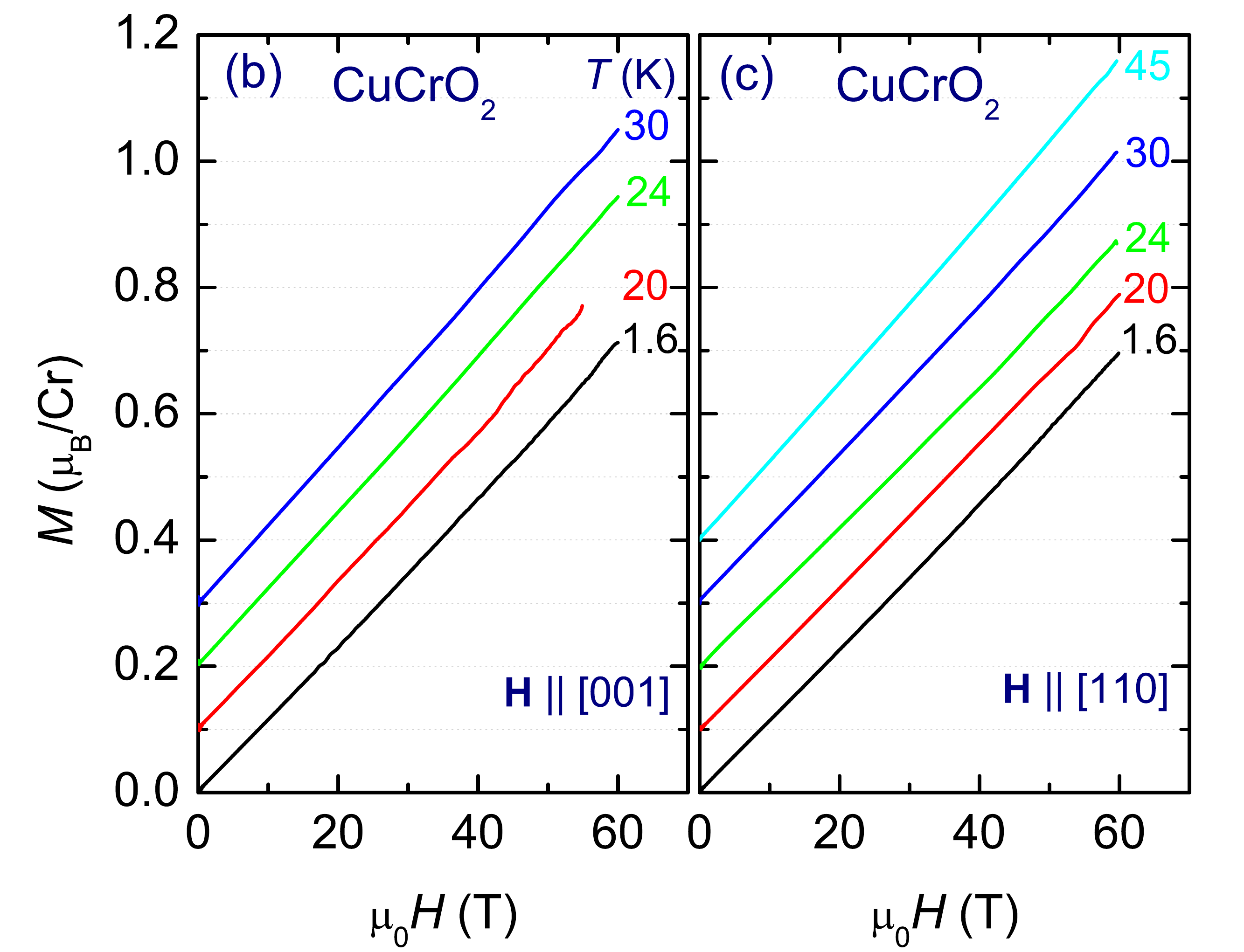}
\caption{(a) Magnetization $vs$ magnetic field curves, $M(H)$, taken in the up sweep of pulsed magnetic field for $\textbf{H}$ $\parallel$ [110], [1-10], and [001] at $T$ = 1.6 K. The
dashed-line is a linear guide to the eye. Left upper inset shows d$M$/d$H$ for $\textbf{H} \parallel$ [110] and [1-10], where symbols and lines are based on data taken in
superconducting and pulsed field magnets, respectively, at $T$ = 2 K. Right lower inset presents the $M(H)$ data for both the up sweep (symbol) and down sweeps (line) of pulsed
magnetic fields up to 60 T along $\textbf{H}$ $\parallel$ [110] at $T$ = 0.6 K. (b) and (c) $M(H)$ curves for $\textbf{H}$ $\parallel$ [001] and $\textbf{H}$ $\parallel$ [110] at
selected temperatures. For clarity, the curves are shifted from each other by 0.1 $\mu_{B}$/Cr.}
\label{Fig3}%
\end{figure*}

Magnetization of CuCrO$_{2}$ in DC and pulsed fields are shown in Fig. \ref{Fig3}. In Fig. \ref{Fig3} (a) $M(H)$ to 60 T is plotted at 1.6 K, and is roughly linear to 60 T with no
significant magnetic anisotropy between the three orientations of the magnetic field. No hysteresis is seen either, as shown in the right lower inset where $M(H)$ for up- and
down sweeps of $\textbf{H} \parallel$ [110] is plotted for $T = 0.6$ K. $M$ reaches $\sim$ 0.7 $\mu_{B}$/Cr at 60 T and the expected saturation magnetization of $\sim 3 \mu_B$/Cr
linearly extrapolates to $H = 270$ T. A slight deviation from a linear dependence of $M(H)$ is observed for $\mu_{0}H$ $>$ $\sim$ 48 T along $\textbf{H}$ $\parallel$ [001]. Note that,
for $\textbf{H}$ $\parallel$ [110], the observed magnetization curve is consistent with an earlier measurement at 1.6 K up to 50 T in pulsed magnetic field \cite{Yamaguchi2010}.

In the upper inset of Fig. \ref{Fig3} (a), we show a feature in d$M$/d$H$ at 5.3 T and 2 K for $\textbf{H}$ $\parallel$ [1-10], corresponding to the spin flop at $H_f$ from a PSS to a
cycloidal spiral structure, consistent with earlier results \cite{Kimura2009b}.

The temperature evolution of the $M(H)$ curves for $\textbf{H}$ $\parallel$ [001] and [110] are plotted in Fig. \ref{Fig3} (b) and (c) at selected temperatures. The general behavior of
the $M(H)$ curves remains the same for different temperatures, although $M(H)$ curves show an increase in the slope with increasing temperature as the influence of the
antiferromagnetic exchange interaction weakens.

\subsection{Electric Polarization at 4 K}

\begin{figure*}
\centering
\includegraphics[width=0.5\linewidth]{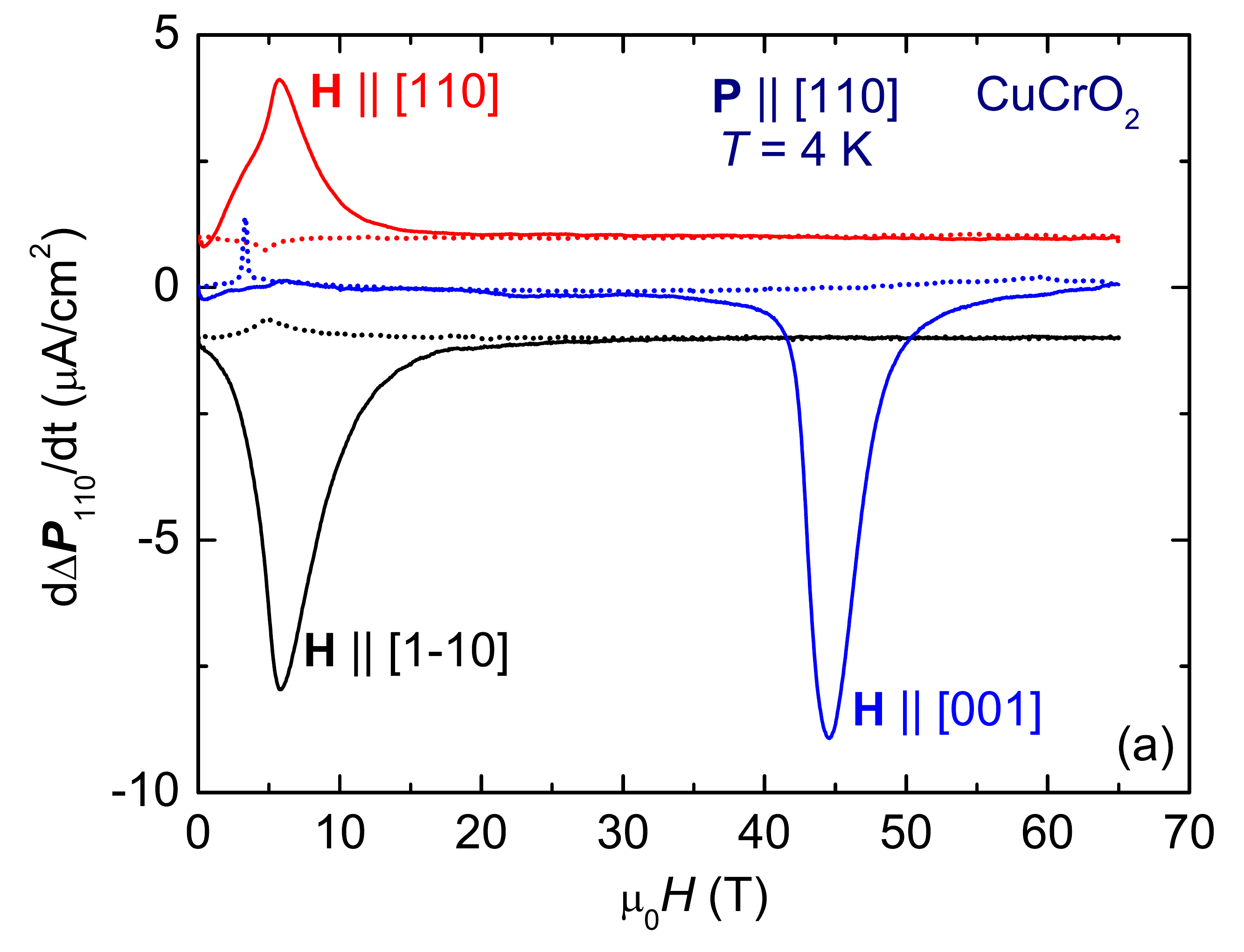}\includegraphics[width=0.5\linewidth]{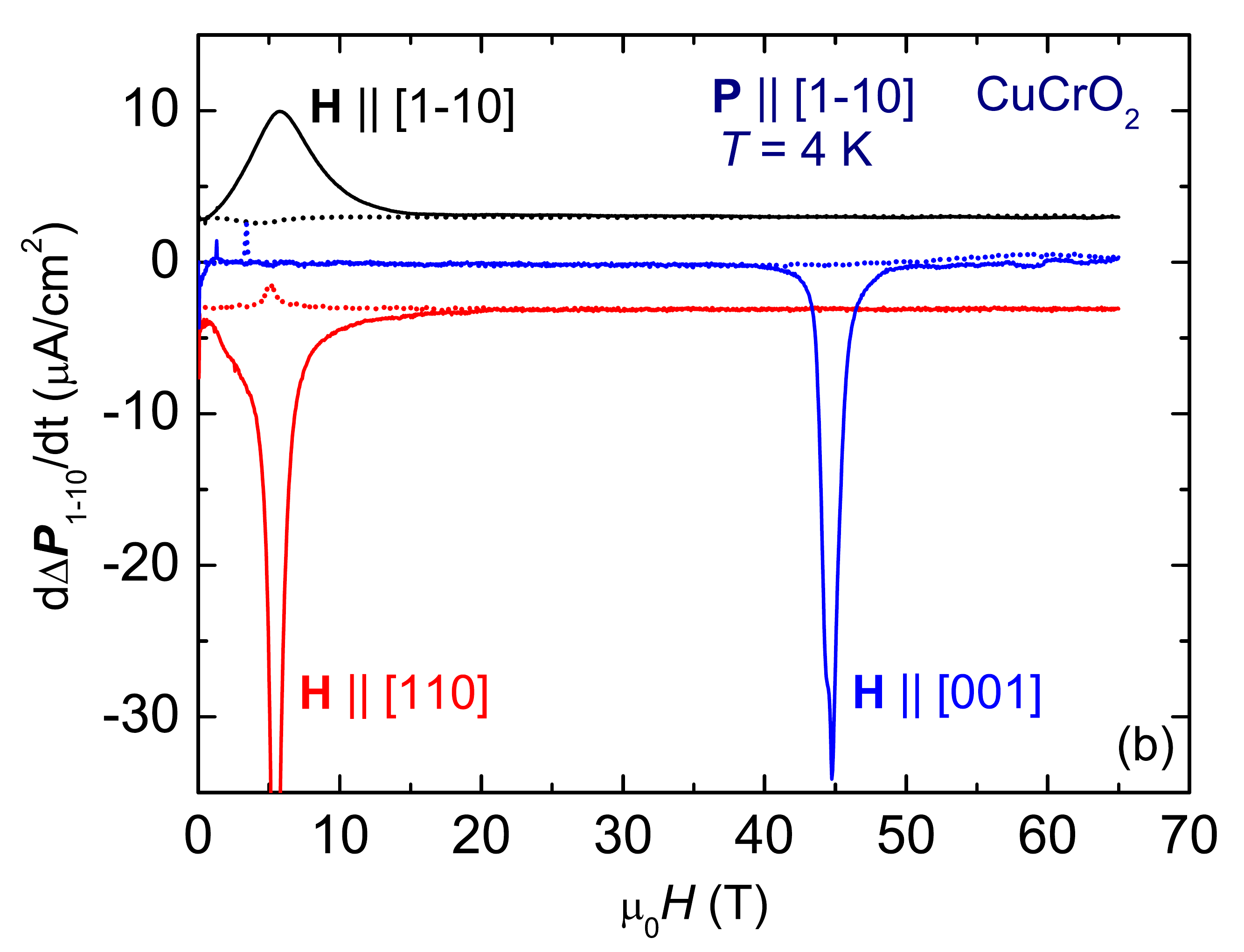}
\includegraphics[width=0.5\linewidth]{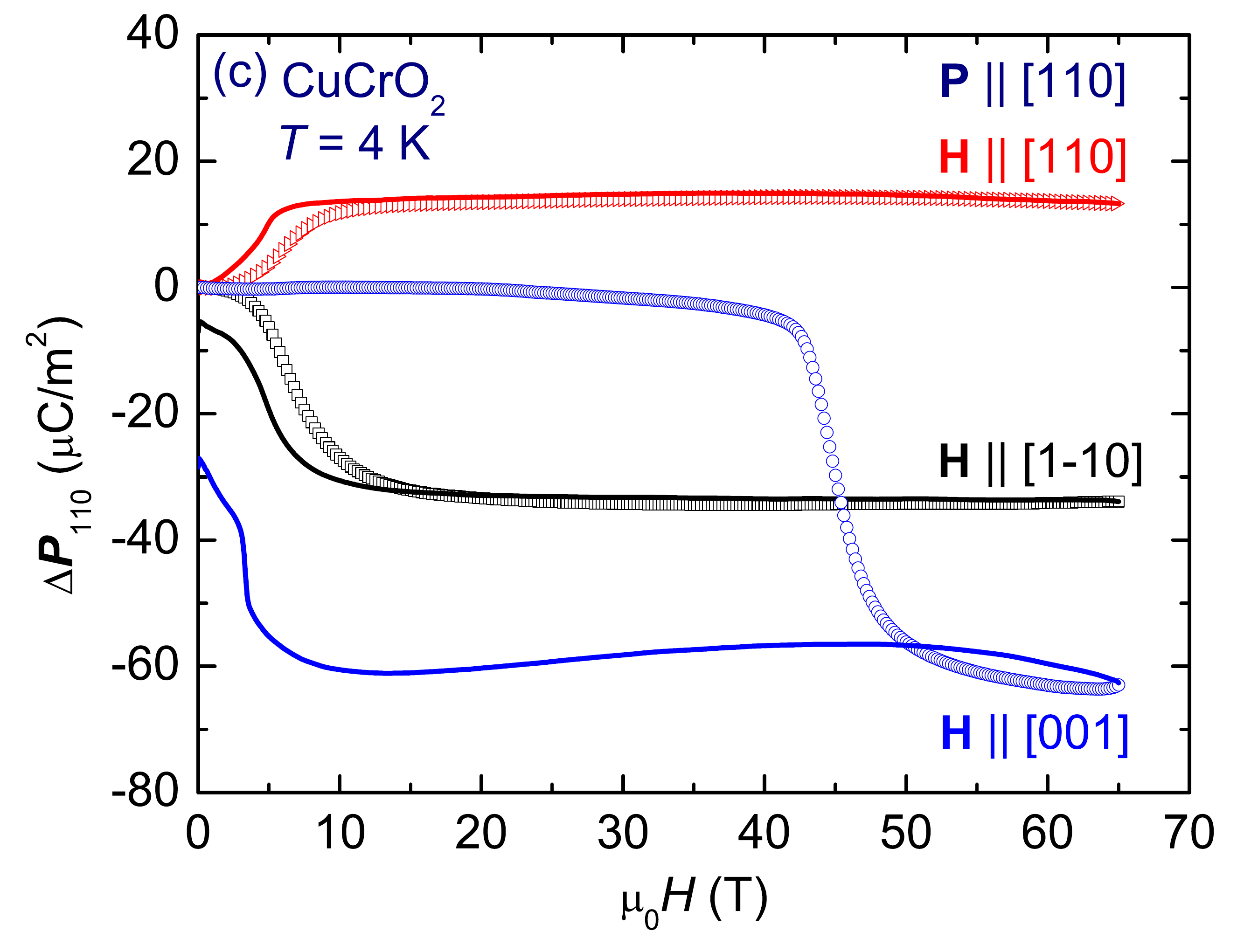}\includegraphics[width=0.5\linewidth]{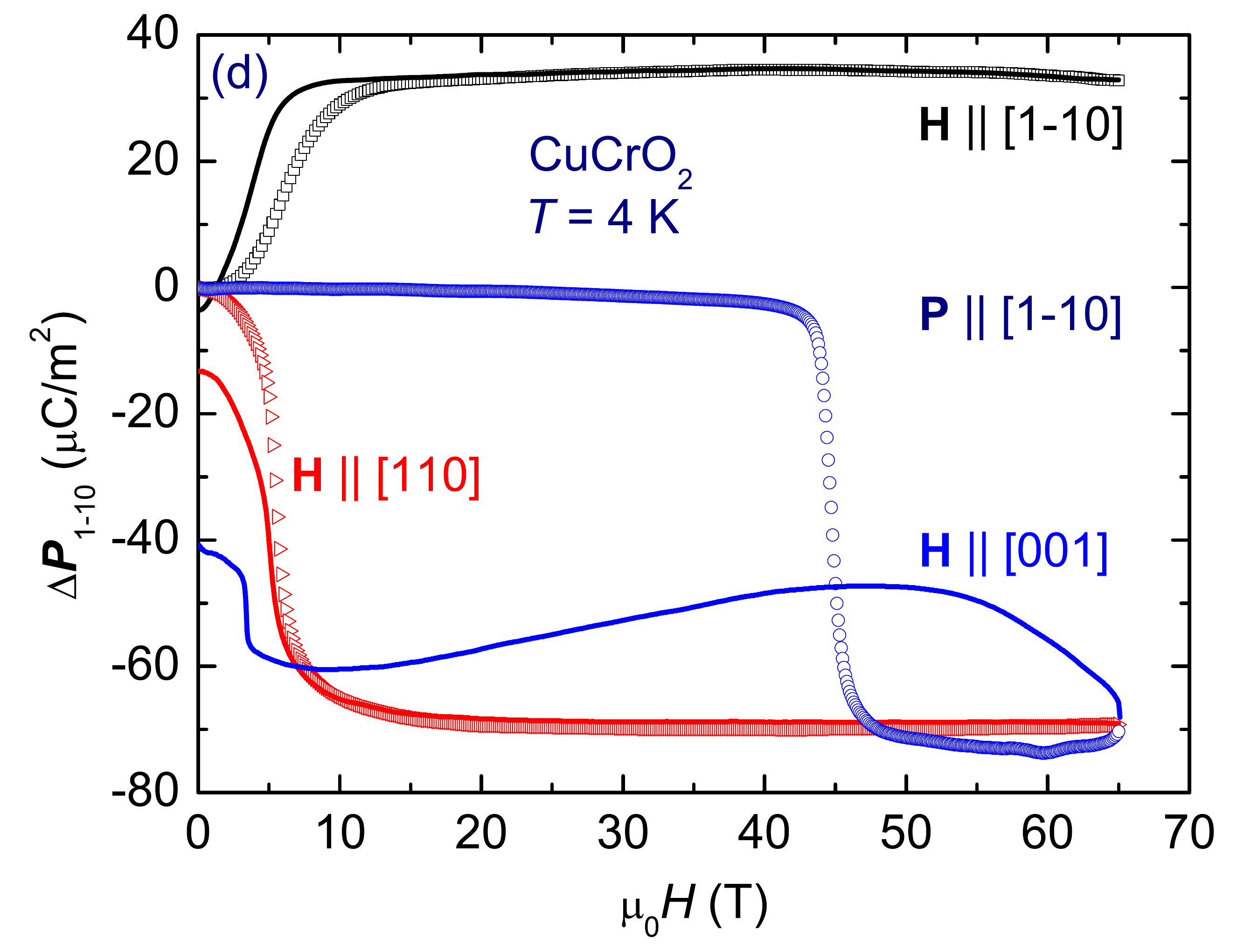}
\caption{Magnetic field variation of the electric polarization $\textbf{P}$ of CuCrO$_{2}$ at $T$ = 4 K for three different orientations of applied magnetic field $\mu_{0}H$. Electric
polarization change with time, d$P$/d$t$, for (a) $\textbf{P}$ $\parallel$ [110] and (b) $\textbf{P}$ $\parallel$ [1-10], plotted as a function of $\mu_{0}H$. Solid and dotted lines
represent up- and down sweeps of magnetic field, respectively. The curves are shifted vertically for clarity. Note that the up sweep of the pulsed magnets is much faster than the
down sweep, resulting in larger signatures in d$\Delta P$/d$t$ for the up sweep than the down sweep for the same $\Delta P$. Electric polarization change $\Delta P(H)$ determined by
integrating d$\Delta P$/d$t$ with respect to time for (c) $\textbf{P}$ $\parallel$ [110] and (d) $\textbf{P}$ $\parallel$ [1-10]. The poling electric field $E_{P}$ was $\sim$ 1.3 MV/m
for $\textbf{P} \parallel$ [110] and $\sim$ 0.6 MV/m for $\textbf{P} \parallel$ [1-10]. Open symbols and solid lines indicate up- and down sweep of magnetic field, respectively.}
\label{Fig4}%
\end{figure*}

\begin{figure}
\centering
\includegraphics[width=1\linewidth]{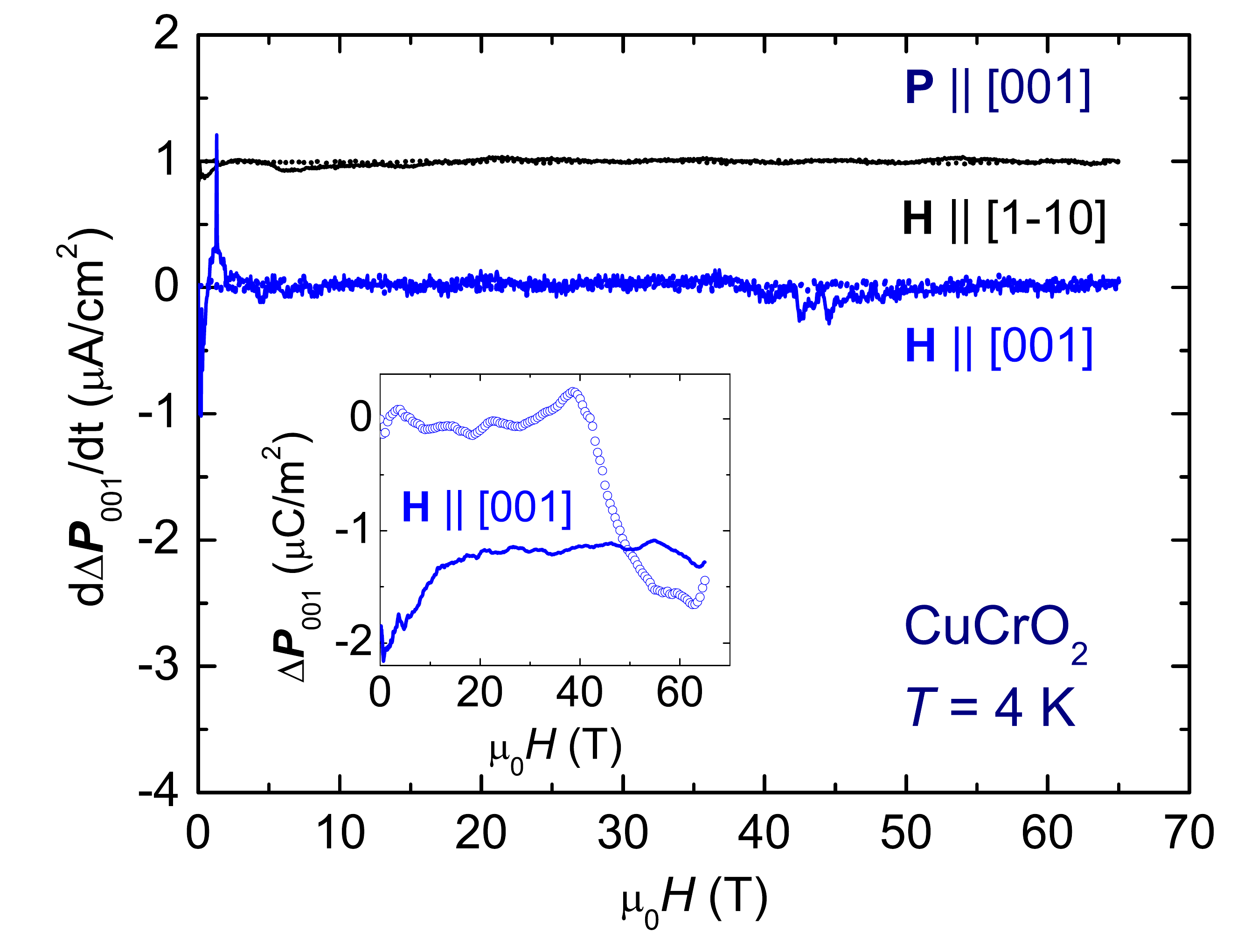}
\caption{Electric polarization change with time, d$\Delta P$/d$t$, for $\textbf{P}$ $\parallel$ [001], plotted as a function of $\mu_{0}H$. The curves are shifted vertically for
clarity. Inset: $\Delta P(H)$ for $\textbf{H} \parallel$ [001].}
\label{Fig5}%
\end{figure}

\begin{figure*}
\centering
\includegraphics[width=0.5\linewidth]{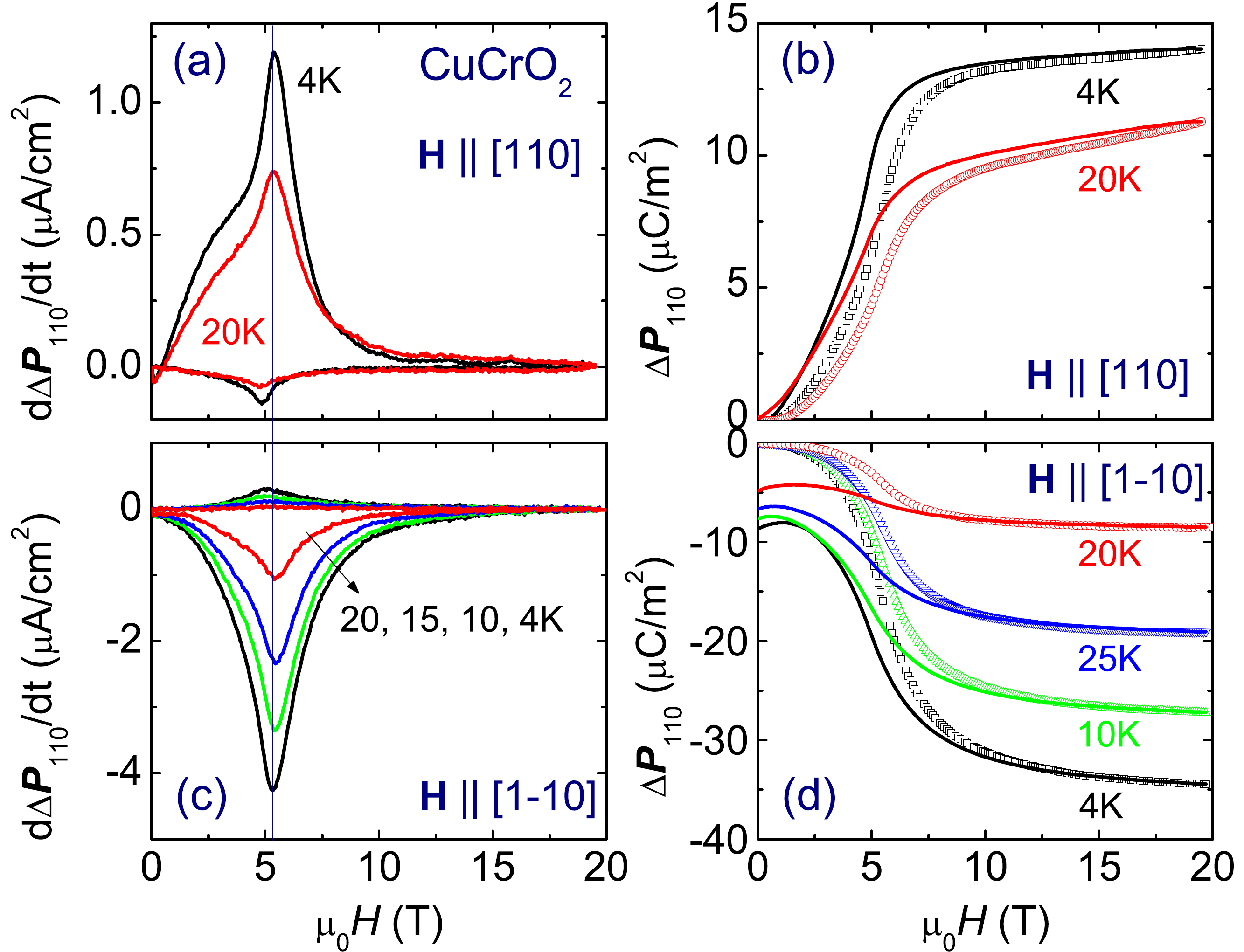}\includegraphics[width=0.5\linewidth]{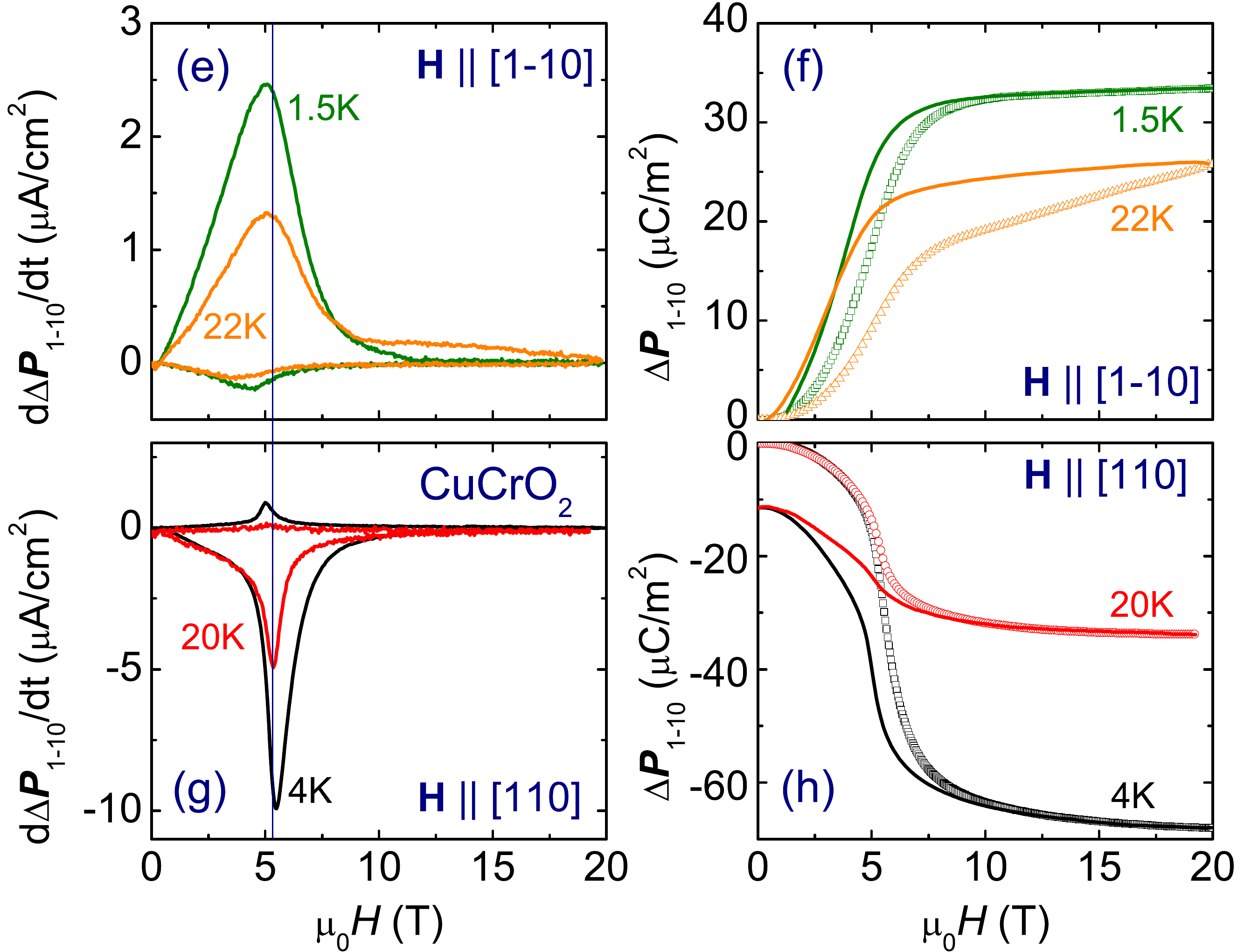}
\caption{Electric polarization change for $\textbf{P} \parallel$ [110] (a-d) and $\textbf{P} \parallel$ [1-10] (e-h) with time, d$\Delta P$/d$t$, plotted as a function of $\mu_{0}H$ at
selected temperatures. (a and e): $\textbf{P} \parallel \textbf{H}$ and (c and g): $\textbf{P} \perp \textbf{H}$. The vertical line is guided to the eye, indicating a transition
$H_{f} \sim 5.3$ T for the up sweep of the magnetic field. (b, d, f, h) $\Delta P(H)$ determined by integrating d$\Delta P$/d$t$ with respect to time. Open symbols and solid lines
indicate up- and down sweep of the magnetic field, respectively.}
\label{Fig6}%
\end{figure*}

In Figs. \ref{Fig4} and \ref{Fig5} we show the measured magnetoelectric current d($\Delta P$)/d$t$ and its integral, the magnetic field-induced change in electric polarization, $\Delta
P(H)$ of CuCrO$_{2}$ at $T$ = 4 K. The temperature evolution of $\Delta P(H)$ is shown in Figs \ref{Fig6}, \ref{Fig7}, and \ref{Fig8}.

When $\textbf{H}$ is applied in the \textbf{ab}-plane in Fig. \ref{Fig4} ($\textbf{H} \parallel$ [1-10] and $\textbf{H} \parallel$ [110]), we observe transitions in $P(H)$ near $\sim$
5.3 T that are consistent with earlier results, and have been attributed to a spin-flop of one magnetic domain from PSS to cycloidal order \cite{Kimura2009b}. This transition occurs
both for $\textbf{P} \parallel$ and $\perp$ to $\textbf{H}$. In these transitions, the electric polarization always becomes larger along the direction of the applied $H$, and smaller
perpendicular to $H$. For magnetic fields above this transition, we find that $\textbf{P}$ remains constant, but nonzero, up to 65 T.

When $\textbf{H}$ is applied along [001] (Fig. \ref{Fig4}), we observe a previously unreported anomaly in $\Delta P(H)$ for $\textbf{P} \parallel$ [110] and [1-10]. The electric
polarization is extraordinarily hysteretic; it is suppressed at 45 T on the up sweep, and then regains most of its original value in a step near 3 T on the down sweep. Similar behavior
was observed in three different samples.

For \textbf{P} $\parallel$ [001], $\Delta P(H)$ for both $\textbf{P} \parallel \textbf{H}$ and $\textbf{P} \perp \textbf{H}$  is insensitive to the entire magnetic field range measured
as shown in Fig. \ref{Fig5}. This observation is consistent with earlier reports \cite{Kimura2008, Kimura2009b} that in zero magnetic field the electric polarization is absent along
\textbf{P} $\parallel$ [001]. We observed a very small change of $\Delta P(H)$ for $\textbf{H}$ $\parallel$ [001] near $\sim$ 1.5 and $\sim$ 40 T, as shown in the inset of Fig.
\ref{Fig5}. However, the variation is an order of magnitude smaller than that for other orientations of $\textbf{P}$. We are assuming that the $\textbf{P} \parallel$ [110] and
$\textbf{P}$ $\parallel$ [1-10] components are intrinsic whereas the small $\textbf{P} \parallel$ [001] component is due to a slight misalignment of the electrode. Therefore, the
electric polarization vectors in CuCrO$_{2}$ are lying within \textbf{ab}-plane in applied magnetic field at least up to 65 T. Details about the variation of $\Delta P(H)$ as a
function of temperature for $\textbf{P} \parallel$ [110] and $\textbf{P} \parallel$ [1-10] are described next.

\subsubsection{Temperature-dependence of $P(H)$ for $\textbf{H}$ $\parallel$ [110] and [1-10]}

Previous works have discussed the electric polarization flop from \textbf{P} $\parallel$ [110] to \textbf{P} $\parallel$ [1-10] by application of magnetic field $H_f = 5.3$ T along
$\textbf{H}$ $\parallel$ [1-10]. However, details about the temperature variation have not been reported. We find that $H_f$ is insensitive to temperature from 1.6 K all the way to
$T_{MF} = 23.6$ K, and in our pulsed field measurements $H_f$ is also insensitive to the direction of magnetic field in \textbf{ab}-plane. Figures \ref{Fig6} (a-d) and (e-h) show the
$\Delta P(H)$ curves for \textbf{P} $\parallel$ [110] and \textbf{P} $\parallel$ [1-10], respectively, at selected temperatures. Although the absolute value of the $\Delta P(H)$ is
suppressed with increasing temperature, the peak position in d$\Delta P$/d$t$ remains the same for all measured temperatures.

The observed $\Delta P(H)$ curves show a large hysteresis, which is enhanced as the temperature is lowered, and has been attributed to the first-order nature of the phase transition
\cite{Kimura2009b}. Interestingly, for the down sweep of magnetic field, the $\Delta P(H)$ curves come back to zero for $\textbf{P} \parallel \textbf{H}$, whereas the $\Delta P(H)$
curves do not come back to zero for $\textbf{P} \perp \textbf{H}$. Note that this behavior with a large hysteresis loop is absent for the electric polarization measurements in
superconducting magnets \cite{Kimura2009b}, where the $\Delta P(H)$ curves are back to zero with narrow region of hysteresis for both configurations of magnetic field. The observed
behavior in pulsed magnetic field might be related to the domain dynamics that can be affected by magnetic field sweep rate.

\subsubsection{Temperature-dependence of $P(H)$ for $\textbf{H}$ $\parallel$ [001]}

\begin{figure}
\centering
\includegraphics[width=1\linewidth]{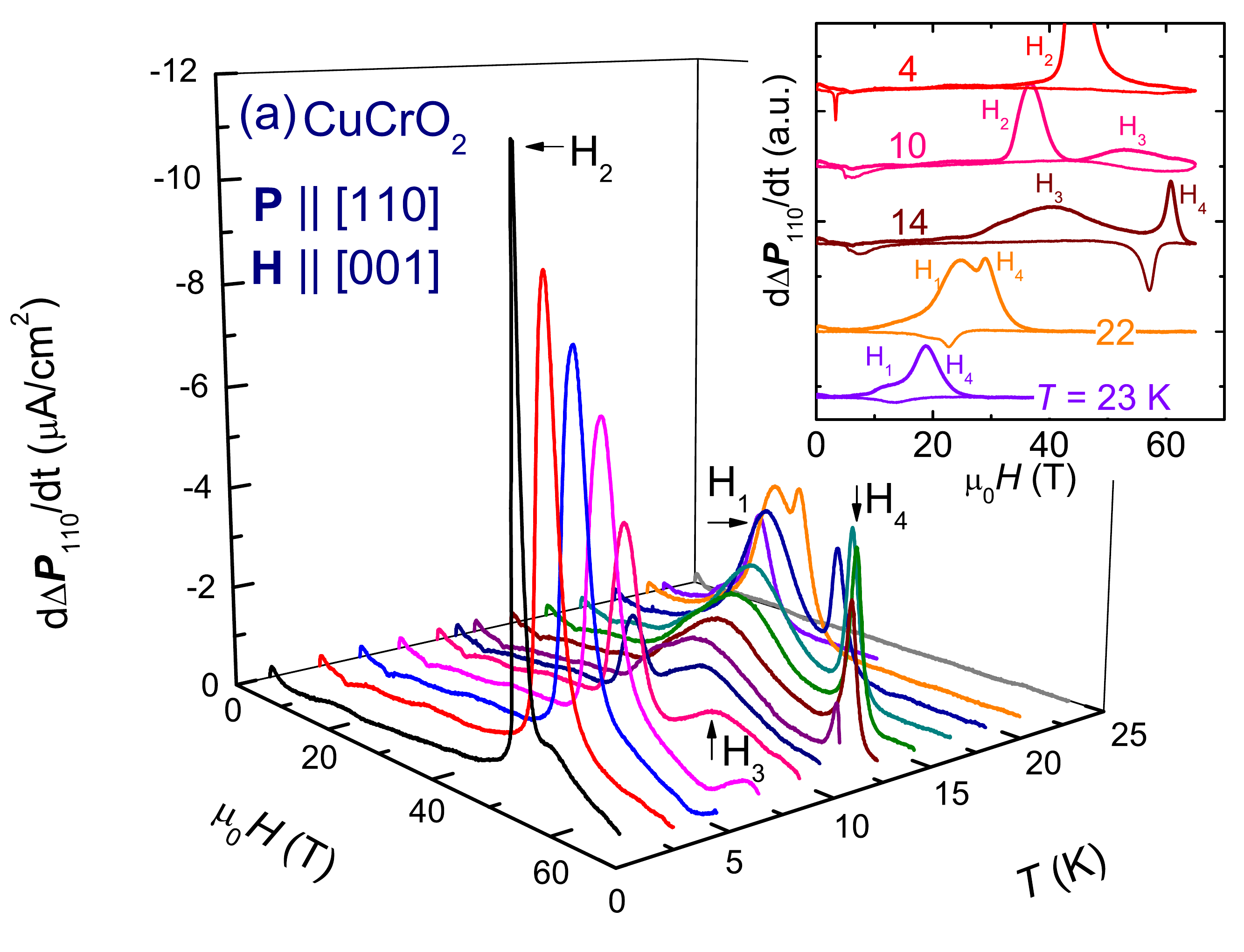}
\includegraphics[width=1\linewidth]{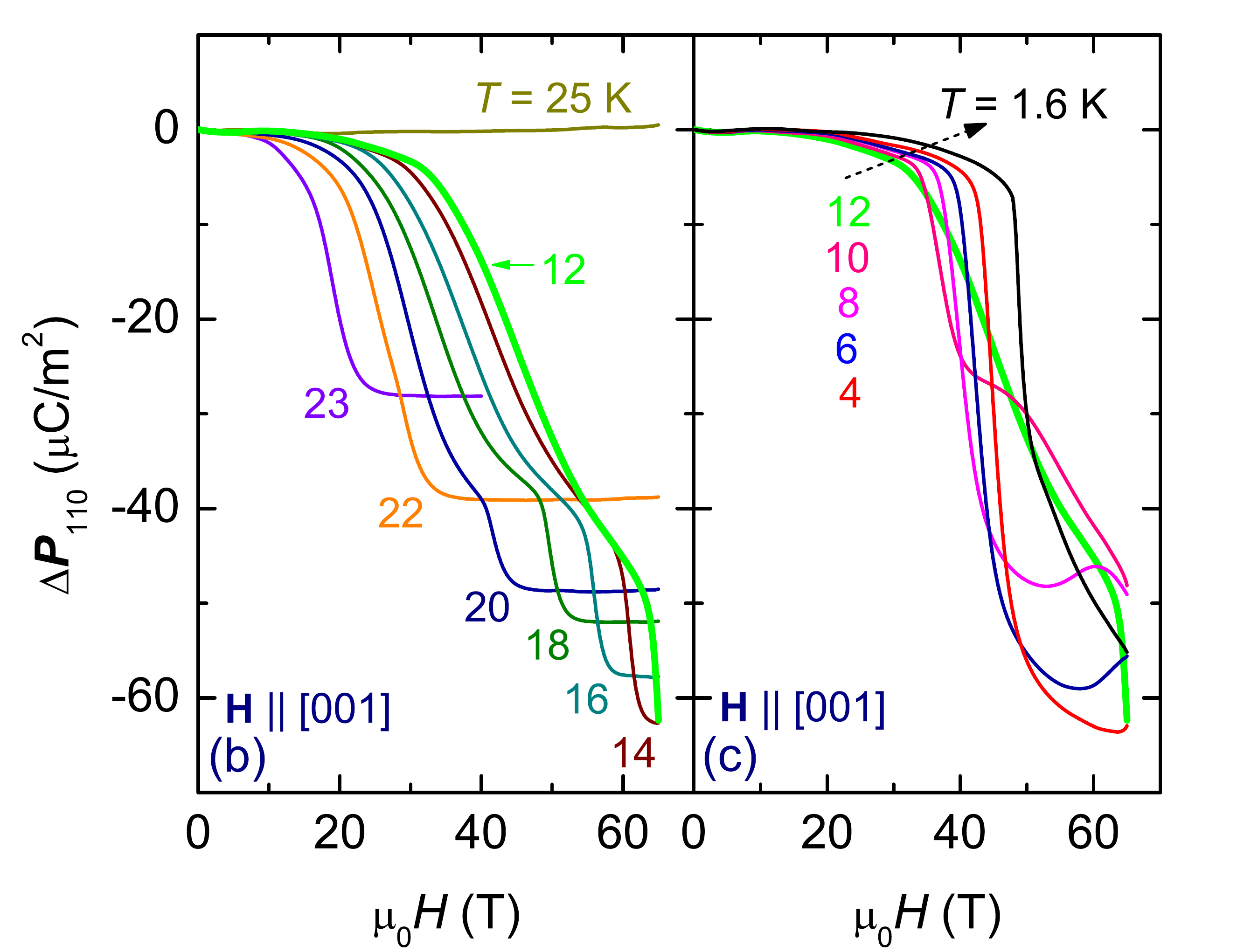}
\caption{(a) 3-D plot of d$\Delta P$/d$t$ for $\textbf{P} \parallel$ [110] along $\textbf{H} \parallel$ [001], plotted as a function of $\mu_{0}H$ and $T$. For clarity, only curves
for up sweep of magnetic fields are plotted. Inset shows the d$P(H)$/d$t$ at $T$ = 23, 22, 14, 10, and 4 K for both up- (thick lines) and down sweep (thin lines) of magnetic field. (b)
and (c) $\Delta P(H)$ curves for up sweep of magnetic field at selected temperatures. For comparison, $\Delta P(H)$ at $T$ = 12 K is plotted in both figures.}
\label{Fig7}%
\end{figure}

\begin{figure}
\centering
\includegraphics[width=1\linewidth]{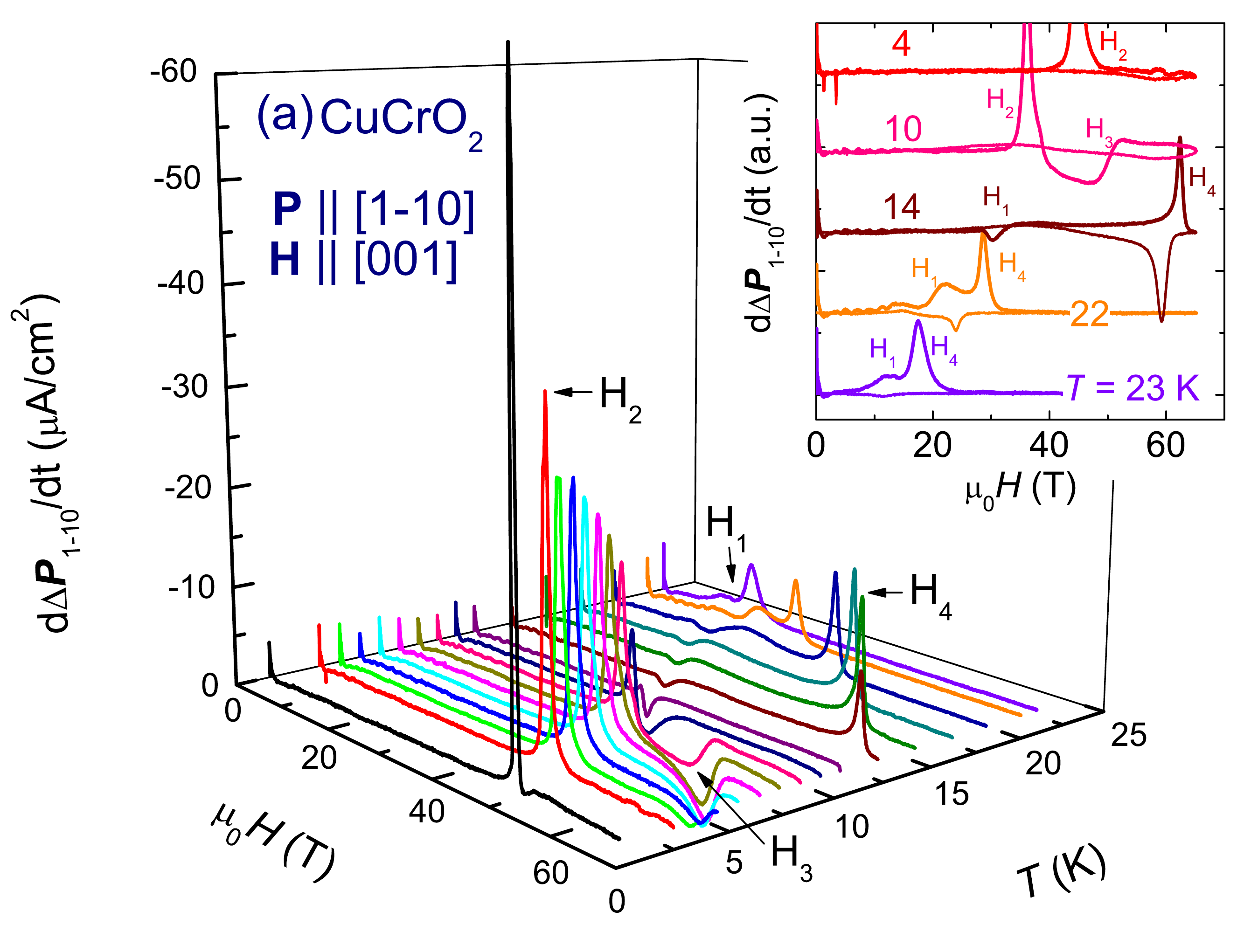}
\includegraphics[width=1\linewidth]{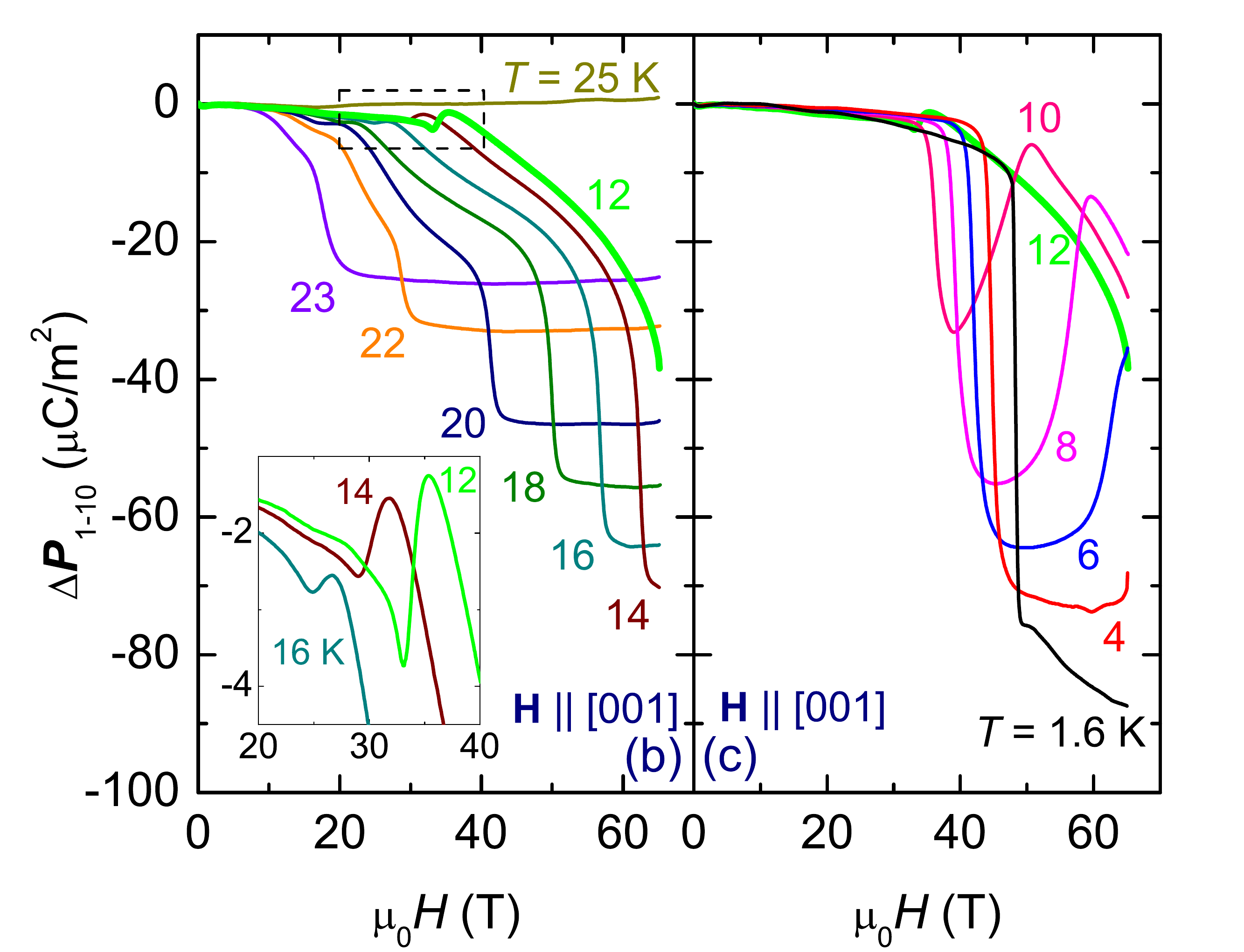}
\caption{(a) 3-D plot of d$P$/d$t$ for $\textbf{P} \parallel$ [1-10] along $\textbf{H} \parallel$ [001], plotted as a function of $\mu_{0}H$ and $T$. For clarity, only curves for the
up sweep of the magnetic field are plotted. Inset shows d$P(H)$/d$t$ at $T$ = 23, 22, 14, 10, and 4 K for both up- (thick lines) and down sweep (thin lines) of magnetic field. (b) and
(c) $\Delta P(H)$ curves for the up sweep of the magnetic field at selected temperatures. For comparison, $\Delta P(H)$ at $T$ = 12 K is plotted in both figures. Inset (b) shows the expanded
plot for the dashed-square region ($T =$ 16, 14, and 12 K).}
\label{Fig8}%
\end{figure}

Figures \ref{Fig7} and \ref{Fig8} show the temperature evolution of \textbf{P} $\parallel$ [110] and \textbf{P} $\parallel$ [1-10] along $\textbf{H}$ $\parallel$ [001] at selected
temperatures. For \textbf{P} $\parallel$ [110], a 3D plot of d$\Delta P$/d$t$ as a function of $\mu_{0}H$ and $T$ is shown in Fig. \ref{Fig7} (a), and the integrated $\Delta P$ is
shown in Figs. \ref{Fig7} (b) and (c).

At the base temperature there is one transition in the electric polarization on the up sweep at $H_{2, up} = 45$ T, and a transition of almost equal size at $H_{2, down}$ = 3 T on the
down sweep. Note that in the raw d$\Delta P(H)$/d$t$ data, these features appear to have different sizes due to the different magnetic field sweep rates on the up- and down sweep. As
the temperature is increased, $H_2$ shifts to lower magnetic fields, and then additional features at higher fields appear on the up sweep, marked $H_1$, $H_3$ and $H_4$. Presumably
$H_3$ and $H_4$ occur above 65 T at base temperature, and then shift down to $H$ $<$ 65 T as the temperature is increased. As the temperature is raised towards $T_{MF}$, the
hysteresis between $H_{2, up}$ = 45 T and $H_{2, down}$ = 3 T narrows until both become 18 T just below $T_{MF}$. The integrated electric polarization $\Delta P(H)$ in Figs.
\ref{Fig7} (b) and (c) shows step-like features corresponding to the peaks in d$\Delta P$/d$t$, and $\Delta P(H)$ actually becomes non-monotonic in $H$ for $T$ = 4, 6, and 8 K.

For $\textbf{P}$ $\parallel$ [1-10], d$\Delta P$/d$t$ and $\Delta P(H)$ are very similar to the previously-discussed case for $\textbf{P}$ $\parallel$ [110], also showing these
additional anomalies at high temperatures (see Fig. \ref{Fig8}). One difference is that the $H_{3}$ peak is negative for $\textbf{P}$ $\parallel$ [1-10], whereas it is positive for
$\textbf{P}$ $\parallel$ [110]. The resultant $\Delta P(H)$ is extremely non-monotonic between 4 and 20 K.

\section{Phase Diagrams and Discussion}

\begin{figure}
\centering
\includegraphics[width=1\linewidth]{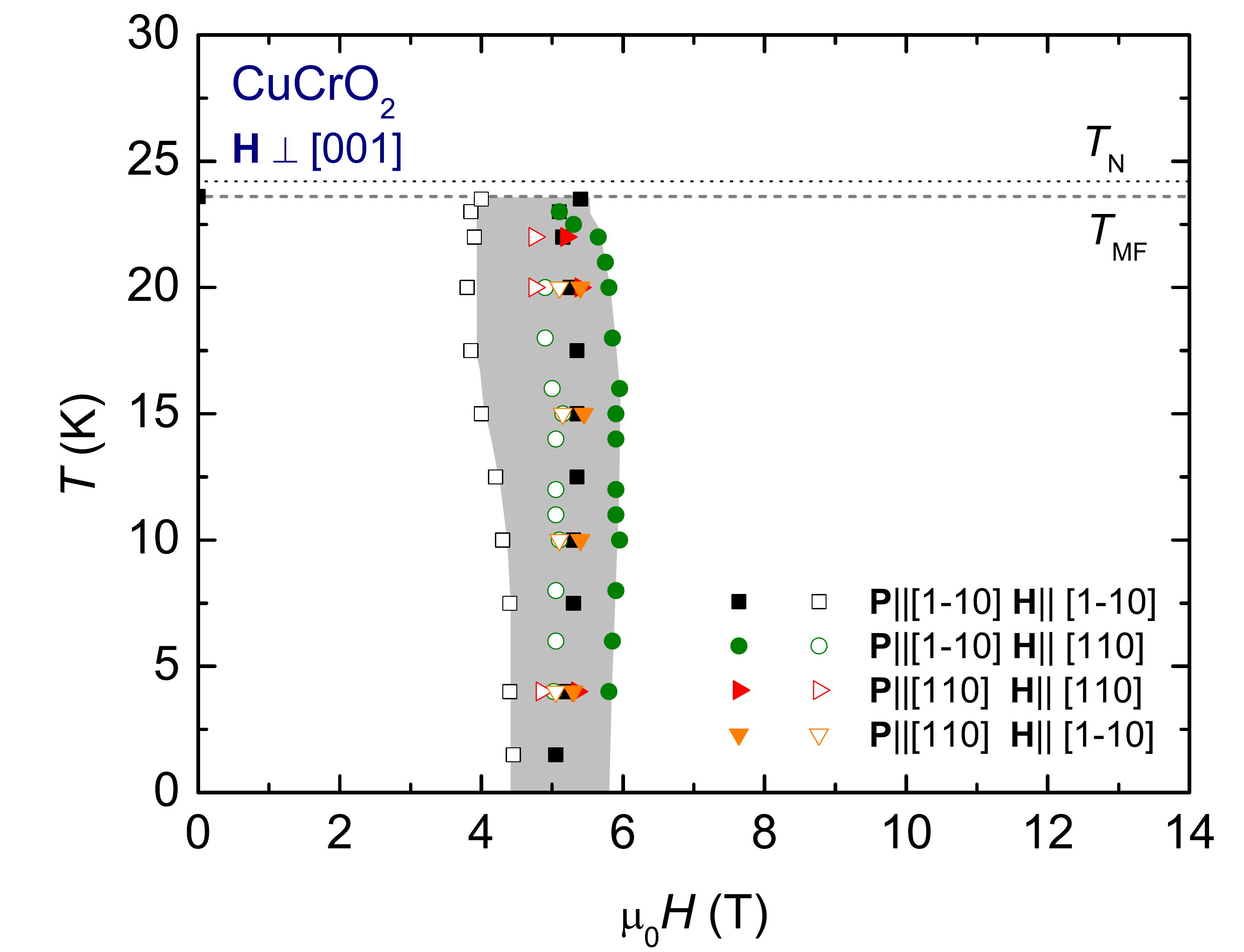}
\caption{Magnetic field-temperature $H-T$ phase diagram of CuCrO$_{2}$ for $\textbf{H} \perp$ [001]. Open and closed symbols represent transitions observed in up- and down sweep of
magnetic field, respectively. The grey area is guided to eyes. Star symbol indicates previously-reported transition in DC fields. The horizontal dotted- and dashed-line indicate the
magnetic transition temperatures $T_{N}$ and $T_{MF}$ at zero magnetic field, respectively.}
\label{Fig9}%
\end{figure}

Our $H-T$ phase diagram of CuCrO$_{2}$ for $\textbf{H}$ in the \textbf{ab}-plane is shown in Fig. \ref{Fig9}. The phase boundary was determined from the peaks in d$P(H)$/d$H$. For $H
= 0$, the system becomes a correlated paramagnet below $T_{N}$ = 24.2 K, and forms a 3D magnetic spiral structure with ferroelectricity below $T_{MF}$ = 23.6 K \cite{Frontzek2012}.
There is no $\textbf{P}$ along the \textbf{c}-axis. For both directions of $\textbf{P}$ in the \textbf{ab}-plane, $\textbf{P}$ is enhanced when $\textbf{P}
\parallel \textbf{H}$ and suppressed when $\textbf{P} \perp \textbf{H}$. The previously-reported PSS to cycloidal spiral phase transition, $H_{f}$, observed in DC measurements near
$\textbf{H} = 5.3$ T when $\textbf{H} \perp \textbf{P}$ in the \textbf{ab}-plane \cite{Kimura2009a} is shown as a star in Fig. \ref{Fig9}. Our pulsed-field data also show features at
$H_{f}$ and we find that $H_{f}$ is independent of temperature up to $T_{MF}$. Although the location of the $H_{f}$ phase transition is consistent between pulsed and DC magnetic field
experiments, the shape of $P(H)$ at the $H_{f}$ phase transition differs. For DC magnetic field measurements, a very sharp first-order transition at $H_{f}$ is observed in $P(H)$ and
dielectric constant, $\epsilon(H)$, along $\textbf{H}$ $\parallel$ [1-10], whereas a gradual transitions in $P(H)$ and $\epsilon(H)$ are observed for $\textbf{H} \parallel$ [110]. This
behavior was explained by gradual domain rearrangement \cite{Kimura2009b}. In pulsed magnetic field measurements, $P(H)$ varies more gradually up to $\sim$10 T for all configurations
we measured. In addition, $\Delta P(H)$ exhibits a very large hysteresis loop compared to DC magnetic field measurements, and does not return to its zero-field value for $\textbf{P}
\perp \textbf{H}$. This discrepancy is probably due to the dynamics of the domain structure in CuCrO$_{2}$. However, this effect does not change our main results. Since $H_{f}$ agrees
between DC and pulsed magnetic fields, the essentials of the different ferroelectric phases remain the same. The magnetization in pulsed magnetic fields is similar to the DC magnetic
field measurements as shown in the inset of Fig. \ref{Fig3} (a), where the peak height and width in d$M$/d$H$ are similar between two different measurements. This is likely because
the longitundinal magnetization results from spin canting within all of the domains, and thus is less sensitive to the domain dynamics.

Previous studies of CuCrO$_2$ suggest \cite{Kimura2008, Soda2010} that the polarization flop at $H_{f}$ in CuCrO$_{2}$ can be interpreted as the rearrangement of multiferroic domains.
The present observations can be explained by the same approach. In zero applied electric field, there are six possible ferroelectric domains as shown in Fig. \ref{Fig1} (c). If the
poling electric field \textbf{E} is applied along [110], the three ferroelectric domains closest to [110] are selected. By applying sufficiently high magnetic field, $\mu_{0}H$ $>$
$H_{f}$, all three domains will transform into the one domain parallel to $\textbf{q}$ for $\textbf{H}$ $\parallel$ [110] and into the one domain perpendicular to $\textbf{q}$ for
$\textbf{H}$ $\parallel$ [1-10]. When the poling electric field \textbf{E} is applied along [1-10], two ferroelectric domains are selected. For $\mu_{0}H$ $>$ $H_{f}$, these two
domains will collapse into the single domain parallel to $\textbf{q}$ for $\textbf{H}$ $\parallel$ [110] and into the single domain perpendicular to $\textbf{q}$ for $\textbf{H}$
$\parallel$ [1-10], respectively. Thus, $\textbf{P}$ increases for $\textbf{H} \parallel \textbf{P}$ and decreases for $\textbf{H} \perp \textbf{P}$. In Ref. \cite{Soda2010}, the
neutron scattering experiments have shown that the domain undergoes spin-flop for $\textbf{H} \perp \textbf{P}$, when $\textbf{P} \parallel$ [110] and $\textbf{H} \parallel$ [1-10].

Within the cycloidal phase above $H_f$, $\textbf{P}$ is remarkably insensitive to $H$, despite the linearly-increasing magnetization between 10 and 65 T. However, it is not so
remarkable when we consider that 65 T corresponds to only about a quarter of the saturation magnetization, and also that in most magneto-electric materials, especially those with
magnetostrictive coupling mechanisms, the leading term in the expansion of $P(M)$ is $P \sim M^2$. Since $M$ changes by only $\sim$ 1/6 of it's value between 10 and 65 T, $P$ should
only change by 1/36 of its value. Given that the electric polarization is induced by SIS-breaking magnetic order, we expect that the electric polarization can be suppressed some time
before reaching the full saturation magnetization in a $\sim$ 270 T applied magnetic field.

\begin{figure}
\centering
\includegraphics[width=1\linewidth]{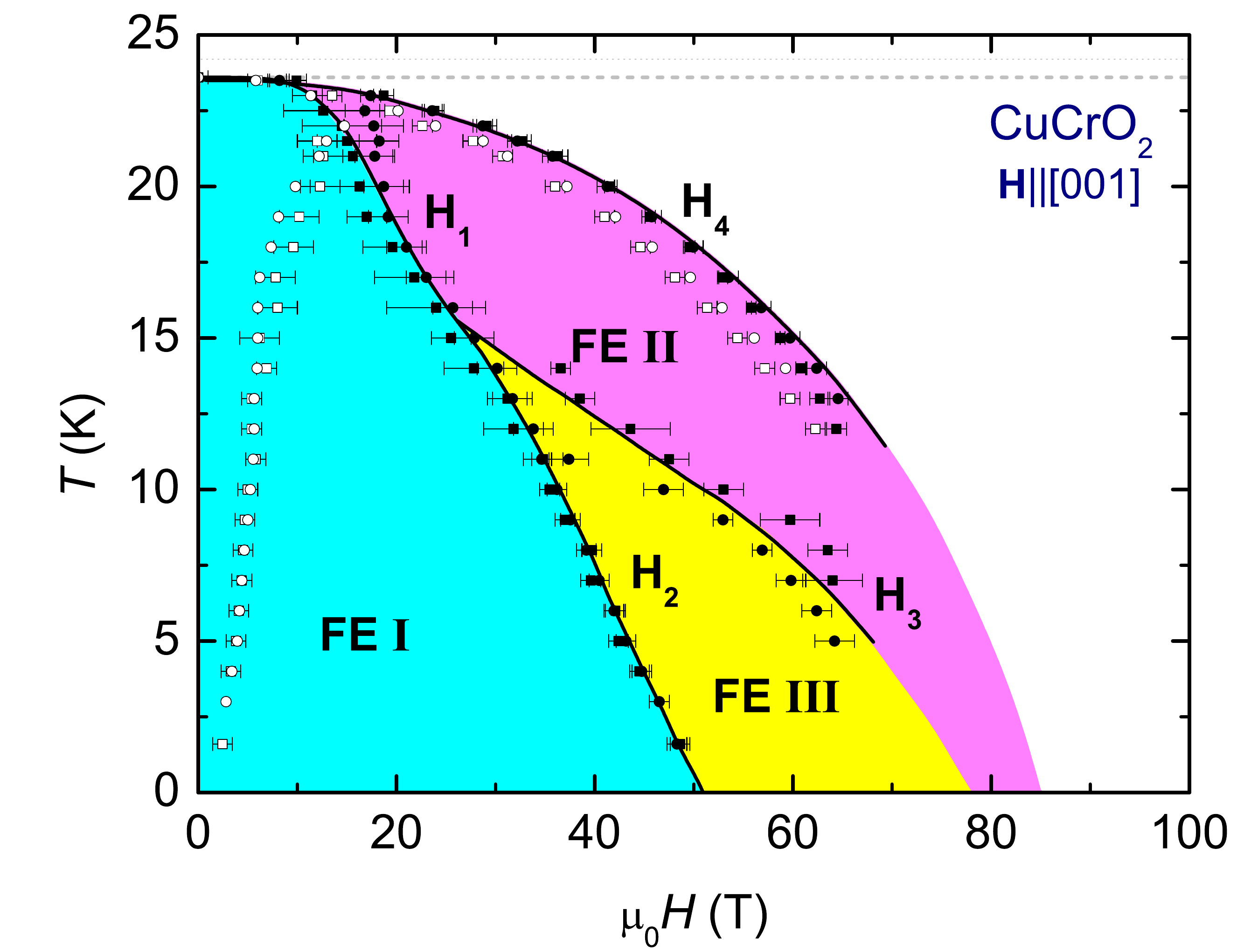}
\caption{Magnetic field-temperature $H-T$ phase diagram of CuCrO$_{2}$ for $\textbf{H} \parallel$ [001]. Solid circles and squares are based on the measurements for $\textbf{P}
\parallel$ [110] and $\textbf{P} \parallel$ [1-10], respectively. Open and closed symbols represent transitions observed in up- and down sweep of magnetic field, respectively. Labels
$H_{1}$, $H_{2}$, $H_{3}$, and $H_{c}$ correspond to the anomalies observed in d$P(H)$/d$t$ (see Fig. \ref{Fig7} and \ref{Fig8}). Thick lines and shaded areas are guided to eyes.}
\label{Fig10}%
\end{figure}

In Fig. \ref{Fig10}, the $H-T$ phase diagram is shown for $H$ along the \textbf{c}-axis obtained from $P(H)$ for \textbf{P} $\parallel$ [110] and \textbf{P} $\parallel$ [1-10]. The
lines are defined by the peak positions in d$P(H)$/d$H$, and different transitions for up- and down sweeps are shown.

The spin-flop transition observed for $H$ in the \textbf{ab}-plane of 5.3 T would also be expected to occur for $H$ along the \textbf{c}-axis, although at a different magnetic field
due to magnetic anisotropy. Thus we can speculate that the 45 T transition ($H_2$) observed for $H$ along the \textbf{c}-axis is the same PSS to cycloidal spiral transition that
occured at 5.3 T for $H$ in the \textbf{ab}-plane. This speculation is also in agreement with the observed hysteresis. For the down sweep of the magnetic field, an increase of the
polarization is observed around 5 T, but the complete initial polarization is not recovered. This can be understood by considering that the initial state was mainly a one-domain state because the sample was prepared by cooling in an electric field. However, on the down sweep of magnetic field, the cycloidal phase decays into a 3-domain PSS state, resulting in less $P$ along the measured direction. 

The assorted transitions observed in $\Delta P(H)$ for different directions of $\textbf{P}$ and $\textbf{H}$ are only weakly observable in the mostly-linear $M(H)$ curves up to 65 T.
In light of previous neutron diffraction measurements, we conclude that the difference between the $M(H)$ and $P(H)$ curves can be related to the fact that $P(H,E)$ results largely
from domain physics of domains that have little net magnetization \cite{Soda2009a,Soda2010} whereas the measured longitudinal $M$ is likely the result of spin canting that occurs in
all domains.

Finally, we compare CuFeO$_{2}$ and CuCrO$_{2}$, both of which have multiferroic properties that have been explained by the mechanism proposed by Arima \cite{Arima2007}. They have the
same crystal structure, where the magnetic Fe and Cr ions are located on triangular lattice planes stacked along \textbf{c}-axis. The MF phase in CuFeO$_{2}$ is observed in an applied
magnetic field between critical fields (7 and 13.5 T) or through doping. By contrast, the CuCrO$_{2}$ is multiferroic at zero magnetic field. In addition, we observed in this study
that the multiferroic phase in CuCrO$_{2}$ can be suppressed by applied magnetic fields $\textbf{H}$ $\parallel$ [001]. Whereas the inter-plane coupling in CuCrO$_{2}$ is
ferromagnetic, the inter-plane coupling in CuFeO$_2$ is antiferromagnetic, and non-frustrated. The nearest neighbor \textit{intra}-layer exchange interaction $J_{1}$ is one order of
magnitude stronger in CuCrO$_{2}$ than CuFeO$_{2}$ \cite{Frontzek2011}. Therefore, the critical field for the suppresion of the multiferroic phase in CuCrO$_{2}$ is much larger than
the critical fields in CuFeO$_{2}$.

A theoretical work \cite{Fishman2011} has examined CuCrO$_2$ using a simplified two dimensional triangular lattice AFM considering intralayer exchange parameters up to 3rd neighbor,
but setting interlayer interactions to zero. This model qualitatively reproduces our phase diagrams, but does not predict the details we observe for $\textbf{H}$ $\parallel$ [001]
(Fig. \ref{Fig10}). In this model, the next stable magnetic state in applied magnetic fields is a three-sublattice phase with $M$ = 1/3 $M_{s}$. At this phase transition, there is
predicted to be a magnetization plateau, and the complete suppression of the electric polarization. If the prediction of this collinear three-sublattice phase is correct, then we can
extrapolate from our magnetization data that the transition should occur at $\sim$ 90 T. Based on the model, the magnetic structure is always in a three-sublattice spiral state for all our measurements up to 65 Telsa. When we interpret the transitions as the spin-flop to the cycloidal state, then the emerging phase between $H_{2}$ and $H_{3}$ is especially
interesting. Here, the magnetic structure is still a PSS, but with a strongly-reduced electric polarization. The applied magnetic field couples in the \textbf{c}-direction to the ferromagnetic
interlayer exchange interaction. The stronger interlayer exchange might lead to new stacking sequence with a \textbf{c}-component of 1/2 in the propagation vectors. According to Ref.
\cite{Rastelli1986} this would subsequently change the propagation vector from (1/3 - $\epsilon$, 1/3 - $\epsilon$, 0) to (1/3 - $\epsilon$, 1/3, 1/2).

\section{Summary}

The $H-T$ phase diagram of single crystalline CuCrO$_{2}$ is explored up to 60 T in magnetization and 65 T in electric polarization for all the principle directions of $\textbf{P}$ and
$\textbf{H}$. At low temperatures, ferroelectricity persists beyond 65 T for all directions of $\textbf{P}$ and $\textbf{H}$ (except for $\textbf{P} \parallel$ [001], that shows no
ferroelectricity for all $\textbf{H}$). We find that the $P(H)$ curves for $\textbf{H} \parallel$ [001] reveal a complex new phase diagram with features on the up sweep of the
magnetic field. We speculate that magnetic fields along the \textbf{c}-axis induce a PSS to cycloidal transition. At lower temperatures a non-monotonic evolution of the electric polarization with field could be consistent with the spiral wave vector stretching
through incommensurate and commensurate wave vectors. On the other hand, for $\textbf{H}$ in the \textbf{ab}-plane, we observe the previously-reported spin flop transition to a
cycloidal spiral magnetic phase near 5.3 T, and find that it is not sensitive to temperatures between 1.6 K and the ordering temperature at $T_{MF} = 23.6$ K. We find that the electric polarization remains essentially constant between $H_{f}$ and the maximum magnetic field of 65 T.

The $M(H)$ curves reveal a quasi-linear isotropic magnetic field dependence up to 60 T, with very weak signature of the spin-flop transition for $\textbf{H}$ along [1 -1 0] at 5.3 T,
and possibly a faint upward deviation from linear behavior at the 45 T transition for $\textbf{H}$ along [001]. The magnetization value at 60 T reaches $\sim$ 0.7 $\mu_{B}$/Cr, which
is far below the saturation value. By a simple linear extrapolation, any 1/3 $M_{s}$ plateaus should occur near $\sim$ 90 T, and complete saturation by 270 T.


CuCrO$_2$ shows strong magnetic frustration due to its triangular spin configuration in the \textbf{ab}-plane, and this is evidenced by the discrepancy between the $\sim 24$ K 2-D ordering
temperature and the $\sim$ 270 T saturation magnetization and $\sim$ 200 K Curie-Weiss temperature. Thus as is typical for frustrated systems we observe multiple magnetic and ferroelectric phases accessibly by
relatively small magnetic fields relative to the saturation magnetic field. However, within most of the phases there is almost no observable variation of the electric polarization
with magnetic field. This may be due to the fact that in most type-two multiferroics, the leading behavior of $P$ is $M^2$, and so $P$ will show most of its magnetic field-dependence at higher magnetic fields that we have applied.

In conclusion, CuCrO$_2$ is a frustrated magnet with a rich variety of magnetic and ferroelectric phases that couple to each other and extend beyond 65 T.

\begin{acknowledgments}
We thank Neil Harrison for the loan of his extraction magnetometer, and valuable discussions with Cristian Batista. Work at the NHMFL was supported by the US National Science Foundation through Cooperative Grant No. DMR-1157490, the State of Florida, and the US Department of Energy. Measurements at LANL were also supported by the Department of Energy's Laboratory Directed Research and Development program. A. P. and G. E. acknowledge funding by the Scientific User Facilities Division, Office of Basic Energy Sciences, US Department of Energy. The research leading to these results has received funding from the European Community's Seventh Framework Programme (FP7/2007-2013) under grant agreement No. 290605 (PSIFELLOW/COFUND). 
\end{acknowledgments}



\clearpage

\clearpage

\end{document}